\documentclass[twocolumn,prl,showpacs,preprintnumbers]{revtex4}
\usepackage{amsmath} 
\usepackage{amsfonts} 
\usepackage{amssymb}
\usepackage{graphicx} 

\newcommand{\half}{\tfrac{1}{2}}

\begin{document}

\preprint{MIT-CTP-3714}
\preprint{KEK-TH-1062}
\preprint{OIQP-05-23}

\title{Hawking Radiation from Charged Black Holes via Gauge and
Gravitational Anomalies} 
\author{Satoshi Iso$^{a,b}$} 
\email{satoshi.iso@kek.jp}
\author{Hiroshi Umetsu$^c$}
\email{hiroshi_umetsu@pref.okayama.jp}
\author{Frank Wilczek$^a$}
\email{wilczek@mit.edu}  

\affiliation{
$^a$ Center for Theoretical Physics, Laboratory for Nuclear
Science and Department of Physics, Massachusetts Institute of Technology,
Cambridge, Massachusetts 02139, USA \\
$^b$ Institute of Particle and Nuclear Studies, High Energy Accelerator
Research Organization(KEK),  
Oho 1-1, Tsukuba, Ibaraki 305-0801, Japan \\
$^c$ Okayama Institute for Quantum Physics,
 1-9-1 Kyoyama, Okayama 700-0015, Japan
}

\begin{abstract}

Extending \cite{Robinson:2005pd}, 
we show that in order to
avoid a breakdown of general covariance and gauge invariance at the quantum
level the total flux of charge and energy in each outgoing partial wave of a
charged quantum field in a Reissner-Nordstr\"om black hole background must be
equal to that of a $(1+1)$ dimensional blackbody at the Hawking temperature 
with the appropriate chemical potential.   
\end{abstract}

\maketitle

%%%%%%%%%%%%%%%%%%

{\it Introduction}.---
There are several  derivations of Hawking radiation. Hawking's original one 
\cite{Hawking:sw, Hawking:rv}, which
calculates the Bogoliubov coefficients between in and out states for 
a body collapsing to form a black holes, is the most direct. 
An elegant derivation based on Euclidean quantum gravity 
\cite{Gibbons:1976ue}   
has been interpreted as a calculation of tunneling through classically 
forbidden trajectories \cite{Parikh:1999mf}.  
It remains of interest to consider alternative derivations, 
since each is based on different assumptions, that might or might not be 
incorporated within a complete theory of quantum gravity.  

Recently Robinson and Wilczek proposed a new partial derivation of 
Hawking radiation \cite{Robinson:2005pd}, which 
ties its existence to the cancellation of gravitational anomalies at the
horizon.  As explained there, this derivation has important advantages: 
it localizes the source of the anomaly at the horizon, where the geometry 
is non-singular yet the equations simplify; it ties in with the effective
field theory approach to the membrane paradigm 
\cite{Thorne:1986iy, Parikh:1997ma}; and the validity of anomalies seems a
particularly reliable assumption, 
since anomalies are a profound feature of quantum field theory.
In this letter we extend the analysis of \cite{Robinson:2005pd} to Hawking 
radiation of charged particles from  Reissner-Nordstr\"om (RN) black holes.  
To do this we will need to consider gauge anomalies at the horizon in
addition to gravitational anomalies.

To identify the ground state of a quantum field, one normally associates
positive energy with occupation of modes of positive frequency. But in
defining positive frequency, one must refer to a specific definition of
time.  
In the Boulware state, vacuum is defined in terms of the Schwarzschild time.  
That is a natural definition of time in the exterior region, 
but it becomes problematic at the horizon.   
In this state freely-falling observers feel a singular flux of 
energy-momentum tensor as they pass through the horizon.  
That unphysical behavior arises from non-trivial occupation of modes 
that propagate nearly along the horizon at high frequency. 
The Unruh vacuum \cite{Unruh:db}, in contrast, uses a time variable adapted 
to the near-horizon geometry to define energy.  
It associates (large) positive energy to the offending modes, so they
are unoccupied.  The basic idea in \cite{Robinson:2005pd} is to form
an effective theory that integrates out the offending modes.  
Having excluded propagation along one light-like
direction, the effective near-horizon quantum field then becomes chiral,  and
contains gravitational anomalies. The original underlying theory is generally
covariant, so failure of the effective theory to reflect this symmetry
should be relieved by introducing an extra effect. It was shown that the
energy-momentum flux associated with Hawking radiation emanating 
from the horizon cancels the anomaly at the horizon.

%%%%%%%%%%%%%%%%%%%%%%%%%%%%%%%%
{\it Setting}: Consider the partial wave decomposition of
a charged  complex field in a RN black hole background. It can be either a
charged scalar or a Dirac fermion. The gauge potential ($A_t=-\phi$) and the
metric of the spacetime are given by  
\begin{eqnarray}
&& A = - \frac{Q}{r}dt, \\
 && ds^2 = f(r) dt^2 - \frac{1}{f(r)} dr^2 -
 r^2d\Omega^2_{(d-2)} \label{eq2}.
\end{eqnarray}
$d^2\Omega_{(d-2)}$ is the line element on the $(d-2)$-sphere and $f(r)$ is 
\begin{equation}
f(r)= 1-\frac{2M}{r}+\frac{Q^2}{r^2} = \frac{(r-r_+)(r-r_-)}{r^2}
\end{equation}
 where 
$r_\pm = M \pm \sqrt{M^2-Q^2}$ are the radii of the outer and inner horizons. 
The surface gravity  at the outer horizon is given by
$\kappa\equiv \half(\partial_r f)|_{r_+}$.

Upon transforming to the $r_*$ ``tortoise'' coordinate defined by
$\frac{\partial r}{\partial r_*}=f(r)$ and performing a partial wave
decomposition, one finds that  the effective radial potentials
for partial wave modes of the  field contain the factor $f(r(r_*)) $ and
vanish exponentially fast near the horizon.  
The same applies to mass terms or other interactions. 
Thus physics near
the horizon can be described using an infinite
collection of massless $(1+1)$-dimensional fields, 
each partial wave propagating in a spacetime with 
a metric given by the ``$r-t$'' section of the full spacetime 
metric (\ref{eq2}).   
This is a kind of dimensional reduction from  $d$-dimensions to $d=2$. 
We exploit that simplification.  Two points should be noted. 
First, the effective two dimensional current or energy momentum tensor are
given by integrating the $d$-dimensional ones over a $(d-2)$-sphere. 
In the four-dimensional case, for example, 
$J_{(2)}^\mu = \int r^2 d\Omega^2 J_{(4)}^\mu=4 \pi r^2 J_{(4)}^\mu(r)$.  
Second, when reducing to
$d=2$, the Lagrangian contains an $r^2$ factor, which can not be interpreted as
the two-dimensional metric. 
This factor can be interpreted as a dilaton background coupled to
the charged fields~\cite{Mukhanov,Kummer}. 

Since the horizon is a null
hypersurface, modes interior to the horizon can not affect physics outside
the horizon,  {\it classically}.  
If we formally remove modes to obtain the effective action 
in the exterior region, 
it becomes anomalous with respect to gauge or diffeomorphism symmetries. 
The underlying theory is, of course, invariant.  
Therefore those anomalies must be
cancelled by quantum effects of the modes that were irrelevant classically.
In the following we show that the conditions for anomaly cancellation at the
horizon are met by the Hawking flux of charge and energy-momentum.  
%%%%%%%%%%%%%%%%%%%%%%%%%%%%%

{\it Gauge anomaly}.---First we investigate  the charged current and gauge
anomaly at the horizon. The effective theory outside of the horizon $r_+$ is
defined in the region $r \in [r_+,  \infty]$. If we first omit the ingoing
modes in the region $r\in [r_+, r_+ + \epsilon]$ near the horizon, the gauge
current exhibits an anomaly there.  The consistent form of $d=2$ abelian
anomaly  (for review, see
\cite{Bertlmann:xk}, \cite{Fujikawa:2004cx}) is given by
\begin{equation}
\nabla_\mu J^\mu = \pm \frac{ e^2}{4 \pi \sqrt{-g}} \epsilon^{\mu \nu}
 \partial_\mu A_\nu, 
\end{equation}
where $+(-)$ corresponds to
left(right)-handed fields respectively and $\epsilon^{01}=1.$ The consistent
anomaly satisfies the Wess-Zumino condition but the current $J^\mu$ 
transforms non-covariantly. 
We can define a new covariant current\cite{Bardeen:1984pm}
\begin{equation}
\tilde{J}^\mu = J^\mu \mp  \frac{e^2}{4 \pi\sqrt{-g}}  A_\lambda
\epsilon^{\lambda \mu}
\end{equation}
 which satisfies  
\begin{equation}
\nabla_\mu \tilde{J}^\mu = \pm  \frac{ e^2}{4 \pi\sqrt{-g}}  \epsilon_{\mu
\nu} F^{\mu \nu}. 
\end{equation} 
The coefficient of the covariant anomaly is twice as
large as that of the consistent anomaly.  

The current is conserved $\partial_r J_{(o)}^r =0$ outside the horizon.  
In the region near the horizon, since there are only outgoing
(right handed) fields, the current satisfies the anomalous equation 
\begin{equation}
\partial_r J_{(H)}^r = \frac{e^2}{4 \pi} \partial_r A_t.
\end{equation}
Hence we can solve them in each region as
\begin{eqnarray}
J_{(o)}^r &=& c_o, \\
J_{(H)}^r &=& c_H + \frac{e^2}{4\pi}
\left( A_t(r) -A_t(r_+) 
\right),
\end{eqnarray}
where $c_o$ and $c_H$ are integration constants.

Under gauge transformations, variation of the effective action 
(without the omitted ingoing modes near the horizon) is given by 
$-\delta W=\int d^2 x \sqrt{-g_{(2)}} \lambda \nabla_{\mu} J_{(2)}^{\mu}$, 
where $\lambda$ is a gauge parameter.
The current is written as a sum of two regions
$
J^{\mu} = J_{(o)}^{\mu} \Theta_+(r) +  J_{(H)}^{\mu} H(r),
$
where $\Theta_+(r) = \Theta(r-r_+-\epsilon)$ and $H(r)=1-\Theta_+(r)$.  
Then by using the anomaly equation, the variation becomes
\begin{eqnarray}
- \delta W &=& \int d^2 x \lambda 
\left[  
\delta(r-r_+ - \epsilon) 
\left(J_{o}^r - J_H^r + \frac{e^2}{4 \pi}A_t \right) \right.
\nonumber \\
&& \left.
 + \partial_r \left(\frac{e^2}{4\pi}A_t H\right)
\right].
\end{eqnarray}
The total effective action must be gauge invariant and the last term should
be cancelled by quantum effects of the classically irrelevant ingoing
modes. The quantum effect to cancel this term is the Wess-Zumino term
induced by the ingoing modes near the horizon. 
The coefficient of the delta-function should also vanish, which relates the
coefficient of the current in two regions; 
\begin{equation}
c_o = c_H - \frac{e^2}{4\pi} A_t(r_+).
\end{equation}
$c_H$ is the value of the consistent current at the horizon. 
In order to determine the current flow, we need to fix the value
of the current at the horizon. Since the condition should be  gauge
covariant, we impose that the coefficient of the covariant current at the
horizon should vanish. 
Since $\tilde{J^r} = J^r + \frac{e^2}{4\pi} A_t(r) H(r)$, 
that condition determines the value of the charge flux to be
\begin{equation}
c_o = - \frac{e^2}{2 \pi} A_t(r_+) = \frac{e^2 Q}{2 \pi r_+}.
\label{Jflux}
\end{equation}
This agrees with the current flow 
associated with the Hawking thermal (blackbody) radiation 
including chemical potential, as will appear presently.  
%%%%%%%%%%%%%%%%%%%%%%%%%%

{\it Gravitational anomaly}.--- We now discuss the flow of the
energy-momentum tensor.  If we neglect  quantum effects of the
ingoing modes, the effective theory exhibits a gravitational anomaly.  
In 1+1 dimensions the consistent anomaly reads
\cite{Alvarez-Gaume:1983ig,Bertlmann:2000da} 
\begin{equation}
\nabla_\mu T^\mu_\nu = 
 \frac{1}{96\pi\sqrt{-g}}\epsilon^{\beta\delta}
 \partial_\delta\partial_\alpha\Gamma^\alpha_{\nu\beta} = {\cal A}_\nu,
 \label{eq1}
\end{equation}
for right-handed fields.
The covariant anomaly, on the other hand, takes the form
\begin{equation}
\nabla_\mu \tilde{T}^\mu_\nu = -\frac{1}{96\pi \sqrt{-g}}
\epsilon_{\mu \nu} \partial^\mu R =\tilde{{\cal A}}_\nu.
\end{equation}

Since the charged field $\phi$ is in a fixed
background of the electric field and dilaton field $\sigma$, 
the energy-momentum tensor is not conserved even classically.  We  first
derive the appropriate Ward identity. 
Under diffeomorphism transformations $x \rightarrow x'=x-\xi$, metric, gauge
and dilaton fields transform as  
$ \delta g^{\mu\nu}=-(\nabla^\mu \xi^\nu + \nabla^\nu \xi^\mu)$, 
$\delta A_\mu = \xi^\nu\partial_\nu A_\mu + \partial_\mu\xi^\nu A_\nu$ and 
$\delta\sigma =\xi^\mu\partial_\mu \sigma$
and the action for matter fields $S[g_{\mu\nu}, A_{\mu}, \phi,\sigma]$ is
invariant. Hence, if there were no gravitational anomaly, the partition
function 
$Z=\int {\cal D} \phi  \exp(i S)$
would obey
\begin{eqnarray}
 && -i \int d^nx  
  \left[
   \delta g^{\mu\nu}(x)\frac{\delta}{\delta g^{\mu\nu}(x)}
   + \delta A_\mu(x) \frac{\delta}{\delta A_\mu(x)}
   \right. \nonumber \\
 && \left. + \delta \sigma(x) \frac{\delta}{\delta \sigma(x)}
  \right]  Z[g_{\mu\nu}, A_\mu, \sigma] = 0.
\end{eqnarray}
Using the energy-momentum tensor
$ T_{\mu\nu} \equiv \frac{2}{\sqrt{-g}}\frac{\delta S}{\delta g^{\mu\nu}}$ 
and current   
$ J^\mu \equiv \frac{1}{\sqrt{-g}}\frac{\delta S}{\delta A_\mu}, $
the Ward identity becomes
\begin{eqnarray}
\nabla_\mu {T^\mu}_\nu 
 =  F_{\mu\nu} J^\mu  
 + A_\nu \nabla_\mu J^\mu 
 - \frac{\partial_\nu \sigma}{\sqrt{-g}} \frac{\delta S}{\delta \sigma}.
\end{eqnarray}
Here we have used the fact that the dilaton couples to the Lagrangian
itself, and kept the term proportional to the gauge anomaly. 
Adding the gravitational anomaly, the Ward
identity becomes 
\begin{eqnarray}
 \nabla_\mu {T^\mu}_\nu 
  = F_{\mu\nu} J^\mu  
  + A_\nu \nabla_\mu J^\mu 
  - \frac{\partial_\nu \sigma}{\sqrt{-g}} \frac{\delta S}{\delta \sigma}
  + {\cal A}_{\nu}. 
  \label{Ward}
\end{eqnarray}
For a metric of the form (\ref{eq2}), the anomaly is purely time-like
(${\cal A}_r=0$) and ${\cal A}_t$ can be written as 
${\cal A}_t \equiv \partial_r N^r_t$
where  $ N^r_t =( f^{\prime 2}+f f^{\prime\prime})/192\pi.$ 
The covariant anomaly is similarly written as
$\tilde{{\cal A}}_t = \partial_r \tilde{N}^r_t$ where
$\tilde{N}^r_t = (f f'' -(f')^2/2)/96\pi.$
At the horizon, since $f=0$, the coefficients have 
opposite signs. 

We now solve the Ward identity (\ref{Ward}) for the $\nu=t$ component.
Since we are considering a static background, the contribution from the
dilaton background can be dropped. 
In the exterior region without anomalies,  the Ward identity is
\begin{equation}
\partial_r T^r_{t (o)} =  F_{r t }  J^r_{(o)},
\end{equation}
and by using $J_{(o)}^r=c_o$ it is solved as
\begin{equation}
 T^r_{t(o)}=a_o + c_o  A_t(r),
\end{equation}
where $a_o$ is an integration constant.
Since there is a gauge and gravitational anomaly near the horizon, 
the Ward identity  becomes 
\begin{equation}
 \partial_r T^r_{t(H)} 
  =  F_{rt} J_{(H)}^r + A_t \nabla_\mu J_{(H)}^\mu
 + \partial_r N^r_t.
\end{equation}
The first and the second term can be combined to become
$F_{rt} \tilde{J}_{(H)}^r $.
By substituting 
$\tilde{J}_{(H)}^r = c_o + \frac{e^2}{2\pi} A_t(r)$ into this equation, 
$T^r_{t{(H)}}$ can be solved as
\begin{equation}
 T^r_{t{(H)}} = a_H + 
 \int^r_{r_+} dr \partial_r \left(
 c_o A_t + \frac{e^2}{4\pi}A_t^2 + N^r_t
 \right).
 \end{equation}
The energy momentum tensor  combines contributions from these two regions 
$T^\mu_{\nu} = T^\mu_{\nu~(o)}\Theta_+ + T^\mu_{\nu~(H)} H$.
Under the diffeomorphism transformation, the effective action changes as
\begin{eqnarray}
 && \int d^2x \sqrt{-g_{(2)}} ~\xi^t \nabla_\mu T^\mu_{t} 
  \nonumber \\
 &=& \int d^2x ~\xi^t
  \left[c_o \partial_r A_t(r) +
   \partial_r \left( \frac{e^2}{4\pi} A_t^2 + N^r_t \right)
  \right. \nonumber \\
 &+& \left.
      \left(T^r_{t~(o)} - T^r_{t~(H)} 
       + \frac{e^2}{4\pi}A_t^2+N^r_t\right) \delta(r-r_+ -\epsilon) 
     \right].
\end{eqnarray}
The first term is the classical effect of the background electric field for
constant current flow. The second term should be cancelled by the quantum
effect of the incoming modes. The coefficient of the last term should vanish
in order to restore the diffeomorphism covariance at the horizon.  
This relates the coefficients: 
\begin{equation}
 a_o = a_H +\frac{e^2}{4\pi}A_t^2(r_+)- N^r_t(r_+) . 
 \end{equation}
In order to determine $a_o$, we need to fix the value of the energy-momentum
tensor at the horizon. As before, we impose a vanishing
condition for the covariant energy-momentum tensor at the horizon. Since
the covariant energy-momentum tensor is related to the consistent one by 
\begin{equation}
\tilde{T}^r_t = T^r_t +\frac{1}{192\pi} (f f'' -2(f')^2),
\end{equation} 
the condition reads $a_H= \kappa^2/24 \pi = 2N^r_t(r_+)$,  
where $\kappa=2 \pi /\beta$ is the
surface gravity of the black hole.
The total flux of the energy momentum tensor 
is given by
\begin{equation}
a_o=\frac{e^2Q^2}{4\pi r_+^2} +N^r_t(r_+)
=\frac{e^2Q^2}{4\pi r_+^2} +\frac{\pi }{12 \beta^2}.
\label{EMflux}
\end{equation}
%%%%%%%%%%%%%%%%%%%%%%%%%%%%
%%%%%%%%%%%%%%%%%%%%%%%%%%%%

{\it Blackbody radiation}.---Now we compare the results 
(\ref{Jflux}), (\ref{EMflux}) with the fluxes from blackbody radiation
moving in the positive $r$ direction at an inverse temperature $\beta$ with
a chemical potential. The Planck distribution in RN black hole  is given by 
\begin{equation}
 I^{(\pm)}(w) = \frac{1}{e^{\beta(w\pm c)}-1}, \ 
 J^{(\pm)}(w) = \frac{1}{e^{\beta(w\pm c)}+1},
\end{equation}
for bosons and fermions respectively and $c=eQ/r_+$\cite{Hawking:sw}. 
$I^{(-)}$ and $ J^{(-)}$ correspond to the distributions for particles with
charge $e$. 
In the zero temperature limit,  if $(\omega \pm c)$ is positive, those
distributions are suppressed exponentially. But if it is negative, they
become $\mp 1$ for bosons or fermions.  
This result needs further interpretation, especially for bosons. In the
bosonic case the absorption coefficient also becomes negative, leading to
the effect known as superradiance \cite{Gibbons:1975kk}. In the more
straightforward fermionic case, 
occupation numbers for these low frequency modes  become $1$ at zero
temperature,  
which leads to a nonzero flux of radiation even at the extremal case. 

To keep things simple,  
we focus now on the fermion case. 
With these distributions, the flux of current and energy-momentum become 
\begin{eqnarray}
 J^r &=& e \int_0^\infty \frac{dw}{2 \pi}
  \left(J^{-}(w) - J^{(+)}(w)\right)
  = \frac{e^2 Q}{2 \pi r_+}, 
  \label{BB1}
  \\
 T^r_{t} &=& \int_0^\infty \frac{dw}{2\pi}
  ~w\left(J^{-}(w) + J^{(+)}(w)\right)
  \nonumber \\
 & = &\frac{e^2 Q^2}{4 \pi r_+^2} + \frac{\pi }{12 \beta^2}.
  \label{BB2}
\end{eqnarray} 
The results (\ref{Jflux}), (\ref{EMflux}) derived from the anomaly 
cancellation conditions coincide  with these results
(\ref{BB1}), (\ref{BB2}). 
Thus, the thermal flux required by black hole 
thermodynamics is capable of canceling the anomaly. 
The actual emission is obtained by propagating the emission from these
sources through the effective potential due to spatial curvature outside the
horizon.  The resulting radiation observed at infinity is that of a
$d$-dimensional grey body at the Hawking temperature.

%%%%%%%%%%%%%%%%%%%%%%%%

%%%%%%%%%%%%%%%%%%%%%%%%%

{\it Discussion}: 
We have derived the flow of charge and energy-momentum from
charged black hole horizons.  In contrast to the conformal anomaly derivation
\cite{Christensen:jc}, we did not need to determine the current 
or energy-momentum tensor elsewhere. 
This is consistent with the universality of Hawking radiation.
Solving the $\nu=r$ component of the Ward identity (\ref{Ward})
would require detailed information on 
the microscopic Lagrangian; likewise other components like $T^r_r$ are
strongly dependent on such nonuniversal physics.  

  We have assumed that covariant forms of current or energy-momentum
tensor should vanish at the horizon. This is natural since  physical
conditions should be gauge or diffeomorphism invariant. 
But we would like to understand more deeply why we
should use the covariant forms instead of the consistent ones
for boundary conditions at the horizon.  
In \cite{Robinson:2005pd}, which employed different procedures 
(integrating out modes in a sandwich surrounding the horizon), 
the consistent anomaly appeared. 
The same results (\ref{Jflux}), (\ref{EMflux}) can be also derived by
calculating the effective action for two-dimensional free fields 
in the RN black hole background and imposing regularity 
at the horizon \cite{IU}. 
This indicates that the regularity 
is closely related to  the vanishing conditions of covariant currents. 

\begin{acknowledgments}
This work was done while S.I. stayed at CTP, MIT under a financial support
from the Japanese Ministry of education, science and culture. S.I. would
like to thank the members at CTP for their warm hospitality and
stimulations. 
S.I. also thanks D. Freedman, K. Fujikawa, Y. Okawa, S. Robinson and
S. Tanimura for  discussions and comments. 

This work is supported in part by funds provided by the U.S. Department
of Energy (D.O.E.) under cooperative research agreement
DE-FC02-94ER40818. 
\end{acknowledgments}

\end{document}